\documentclass[11pt]{article}%
\usepackage{amsfonts}
\usepackage{amssymb}
\usepackage{graphicx}
\usepackage{setspace}
\usepackage{natbib}
\usepackage{amsmath}%
\usepackage{amsthm}%
\usepackage{palatino}
\usepackage{paralist}
\usepackage[top=8pc,bottom=8pc,left=8pc,right=8pc]{geometry}
\usepackage{hyperref}
\usepackage{enumitem}  
\hypersetup{ colorlinks=true,       
    linkcolor=red,          
    citecolor=blue,        
    filecolor=magenta,      
    urlcolor=cyan           
    }
\usepackage{natbib}
\newcommand{\prox}{ \mathop{\mathrm{prox}} }
\newcommand{\R}{\mathcal{R}}

\newcommand{\defeq}{\mathrel{\mathop:}=}
\newcommand{\eqdist}{\stackrel{D}{=}}

\DeclareMathOperator*{\argmax}{arg\,max} 
\DeclareMathOperator*{\argmin}{arg\,min} 

\title{Weighted Bayesian Bootstrap for Scalable Bayes}

\author{Michael Newton\\
\textit{University of Wisconsin}\\
\textit{Madison}\footnote{Newton is Professor of Biostatistics and Medical Informatics at University of Wisconsin-Madison and his effort was supported in part by NIH grant U54AI117924.  email: newton@biostat.wisc.edu.  Polson is Professor of Econometrics and Statistics
at the Chicago Booth School of Business. email: ngp@chicagobooth.edu. Xu is at the Chicago Booth School of Business. email: jianeng@uchicago.edu.}\\
\and
Nicholas G. Polson\\
\textit{Booth School of Business}\\
\textit{University of Chicago}
\and
Jianeng Xu\\
\textit{Booth School of Business}\\
\textit{University of Chicago}
\\}

\date{First Draft: October 2016\\
This Draft: March 2018}

\begin{document}

\maketitle
\begin{abstract}
\noindent We develop a weighted Bayesian Bootstrap (WBB) for  machine learning and statistics. WBB provides uncertainty quantification by sampling from a high dimensional posterior distribution. WBB is computationally fast and scalable using only off-the-shelf optimization software such as TensorFlow. We provide regularity conditions which apply to a wide range of machine learning and statistical models. We illustrate our methodology in regularized regression, trend filtering and deep learning. Finally, we conclude with directions for future research.
\vspace{0.5pc}

\end{abstract}

\newpage

\section{Introduction}
Weighted Bayesian Bootstrap (WBB) is a simulation-based algorithm for assessing uncertainty in machine learning and statistics. Uncertainty quantification (UQ) is an active area of research, particularly in high-dimensional inference problems. Whilst there are computationally fast and scalable algorithms for training models in a wide variety of contexts, uncertainty assessment is still required. Developing computationally fast scalable algorithms for sampling a posterior distribution is a notoriously hard problem. WBB makes a contribution to this literature by showing how off-the-shelf optimization algorithms, such as convex optimization or stochastic gradient descent (SGD) in TensorFlow can also be used to provide uncertainty quantification. 
 \\

\noindent Our work builds on \cite{newton1994approximate} who provide a weighted likelihood Bootstrap (WLB) method for Bayesian inference. They develop a weighted likelihood Bootstrap algorithm together with the appropriate asymptotic analysis to show that such an algorithm 
provides efficient posterior samples. Their bootstrap procedure exploits the fact that the posterior distribution centered at the maximum likelihood estimate (MLE) has a second order expansion that also depends on the prior
and its derivative. The weighted Bayesian Bootstrap (WBB) calculates a series of posterior modes rather than MLEs. This has the advantage that high dimensional posterior modes are readily available particularly using the regularized estimates are fast to compute from convex optimization methods or stochastic gradient descent (SGD) for neural network architectures such as deep learning. By linking WLB and WBB, with modern-day optimization to calibrate estimate, we provide a framework for uncertainty quantification.\\

\noindent Uncertainty estimates are provided at little to no extra cost. Quantifying uncertainty is typically unavailable in a purely regularization optimization method. Another feature that is straightforward to add is a regularization path across hyper-parameters. This is so much easier than traditional Bayesian to do prior sensitivity analysis where hyper-parameters are hard to assess. Rather we use predictive cross-validation techniques.\\

\noindent The rest of the paper is outlined as follows. Section 2 develops our weighted Bayesian Bootstrap (WBB) algorithm. Section 3 provides an application to high dimensional sparse regression, trend filtering and deep learning. WBB can also be applied to Bayesian tree models (\cite{taddy2015bayesian}). Finally, Section 4 concludes with directions for future research. Areas for future study include Bootstrap filters in state-space models (\cite{gordon1993novel})  and comparison with the resampling-sampling
perspective to sequential Bayesian inference (\cite{lopes2012bayesian}), etc.

\section{Weighted Bayesian Bootstrap}
Let  $y$ an $n$-vector of outcomes, $\theta$ denotes a $d$-dimensional parameter of interest and $A$ a fixed $n \times d$ matrix whose rows are the design points (or ``features'') $a_i^T$
where we index  observations by $i$ and parameters by $j$.  
A large number of machine learning and statistical problems can be expressed in the form
\begin{equation}
\label{eqn:canonicalform}
\begin{aligned}
& \underset{\theta \in \R^d}{\text{minimize}}
& &  l(y| \theta) + \lambda\phi(\theta) \, ,
\end{aligned}
\end{equation}
where $l(y| \theta) = \sum_{i=1}^n \log f ( y_i ; a_i^\top \theta )$ is a measure of fit (or ``empirical risk function'') depending implicitly on $A$ and $y$.
The penalty function or regularization term, $\lambda\phi(\theta) $,  
effects a favorable bias-variance tradeoff.  We allow for the possibility that $\phi(\theta)$ may have points in its domain where it fails to be differentiable.  \\

\noindent Suppose that we observe data, $y = (y_1 , \ldots , y_n ) $ from a model parameterized by $\theta$. For example, we might have a probabilistic model that depends on a parameter, $\theta$, where $p(y|\theta)$ is known as the likelihood function. Equivalently, we can define a measure of fit $l(y| \theta) = \log f(y; \theta) = \log p(y|\theta)$. We will make use of the following
\begin{enumerate}[label=(\roman*)]
\item Let $ \hat{\theta}_n \defeq {\rm arg max}_\theta\; p(y |\theta) $ be the MLE, 
\item Let $\theta^*_n \defeq {\rm arg max}_\theta \; p( \theta|y) $ be the posterior mode,
\item Let $ \bar{\theta}_n \defeq E(\theta | y) $ be the posterior mean.
\end{enumerate}
We now develop a key duality between regularization and posterior bootstrap simulation.

\subsection{Bayesian Regularization Duality}

From the Bayesian perspective, the measure of fit, $l(y|\theta) = - \log f(y; \theta)$, and the penalty function, $\lambda\phi(\theta)$, correspond to the negative logarithms of the likelihood and prior distribution in the hierarchical model
\begin{align*}
f(y; \theta) = p(y | \theta) \propto \exp & \{-  l(y|\theta)\} \; , \quad p(\theta) \propto \exp\{ - \lambda\phi(\theta) \} \\
p( \theta | y ) &  \propto \exp\{- ( l(y|\theta) + \lambda\phi(\theta) ) \}.
\end{align*}

\noindent The prior is not necessarily proper but the posterior, $p(\theta | y) \propto p(y | \theta) p(\theta)$, may still be proper. This provides an equivalence between regularization and Bayesian methods. For example, regression with a least squares log-likelihood subject to a penalty such as an $L^2$-norm (ridge) Gaussian probability model or $ L^1$-norm (lasso) double exponential probability model. We then have
\begin{eqnarray}
&\hat\theta_n =& \underset{\theta \in \Theta}{\argmin}\, l(y|\theta),\\
&\theta^*_n =& \underset{\theta \in \Theta}{\argmin}\,  \{ l(y|\theta) + \lambda\phi(\theta) \}.  \label{posterior mode}
\end{eqnarray}

\noindent Let $\partial$ be the subdifferential operator.  Then a necessary and sufficient condition for $\theta^*$ to minimize $ l(y|\theta) + \lambda\phi(\theta)$  is 
\begin{equation}
\label{eqn:subdiffproxgrad}
0 \in \partial \left\{ l(y|\theta) + \lambda\phi(\theta)\right\} = \nabla l(y|\theta) + \lambda\partial \phi(\theta) 
\end{equation}
the sum of a point and a set.  The optimization literature  characterizes $\theta^*$ as the fixed point of a proximal operator
$\theta^* = \prox_{\gamma \phi}\{ \theta^* - \lambda \nabla f(\theta^*) \} $, see \cite{polson2015mixtures} and \cite*{polson2015proximal} for further discussion. \\

\noindent A general class of natural exponential family models can be expressed in terms of the Bregman divergence of the dual of the cumulant transform.
Let $ \phi $ be the conjugate Legendre transform of $\psi$. Hence $ \psi(\theta) = \sup_\mu \; \left ( \mu^\top \theta - \phi(\mu) \right )$. Then we can write
\begin{align*}
p_\psi ( y | \theta ) & = \exp \left ( y^\top \theta - \psi( \theta) - h_\psi(y) \right )\\
 & = \exp \left \{ \inf_\mu \; \left ( ( y - \mu )^\top \theta - \phi(\mu) \right ) - h_\psi(y) \right \}\\
& = \exp \left ( - D_\phi ( y , \mu(\theta) ) - h_\phi (y) \right )
\end{align*}
where the infimum is attained at $ \mu(\theta) = \phi^\prime (\theta) $ is the mean of the exponential family distribution. 
We rewrite $ h_\psi(y) $ is terms of the correction term and $ h_\phi(y)$. Here there is a duality as $D_\phi$ can be interpreted as a Bregman divergence. \\

\noindent For a wide range of non-smooth objective functions/statistical models, recent regularization methods provide fast, scalable algorithms for calculating estimates of the form (\ref{posterior mode}), which can also be viewed as the posterior mode. Therefore as $\lambda$ varies we obtain a full regularization path as a form of prior sensitivity analysis. \\

\noindent \cite{strawderman2013hierarchical} and  \cite{polson2015proximal} considered scenarios where posterior modes can be used as posterior means from augmented probability models. Moreover, in their original foundation of the Weighted Likelihood Bootstrap (WLB), \cite{newton1994approximate} introduced the concept of the implicit prior. Clearly this is an avenue for future research. 

\subsection{WBB Algorithm}
\noindent We now define the weighted Bayesian Bootstrap (WBB). Following \cite{newton1994approximate}, we construct a randomly weighted posterior distribution denoted by
$$
{\bf w} = (w_1, ..., w_n, w_p),\,
p_{\bf w}(\theta | y) \propto \prod_{i=1}^n p(y_i|\theta)^{w_i} p(\theta)^{w_p}
$$
where the weights $w_p, w_i \sim Exp(1)$ are randomly generated weights. It's equivalent to draw
$w_i = \log(1/U_i)$ where $U_i$'s are i.i.d. Uniform (0,1), which is motivated by the uniform Dirichlet distribution for multinomial data. We have used the fact that for i.i.d. observations, the likelihood can be factorized as $p(y|\theta) = \prod_{i=1}^n p(y_i|\theta)$. This is not crucial for our analysis but is a common assumption. Let $\theta^*_{{\bf w},n}$ denote the mode of this regularized distribution. Again, there is an equivalence 

$$
\theta^*_{{\bf w},n} := \underset{\theta}{\argmax} \; p_{\bf w}(\theta | y) \equiv \underset{\theta}{\argmin} \sum_{i=1}^n w_i l_i(y_i|\theta) + \lambda w_{p} \phi(\theta)
$$
where $l_i(y_i|\theta) = -\log p(y_i|\theta)$ and $ \lambda\phi(\theta) = -\log p(\theta)$. Note that we have a weighted likelihood and a new regularization parameter, $\lambda w_p$.\\

\noindent
\noindent  The crux of our procedure is to create a sample of the weighted posterior modes $\{ \theta_{{\bf w},n}^*\}$ (computationally cheap as each sub-problem can be solved via optimization). Our main result is the following:\\

\noindent \textbf{Algorithm: Weighted Bayesian Bootstrap (WBB)}
\begin{enumerate}
\item Iterate: sample ${\bf{w}} = \{w_1, w_2, ..., w_n, w_p\}$ via exponentials. $w_p, w_i\sim Exp(1)$. 

\item For each ${\bf{w}}$, solve $\theta^*_{{\bf w},n} = \underset{\theta}{\argmin} \sum_{i=1}^n w_i l_i(\theta) +  \lambda w_p \phi(\theta)$. 
\end{enumerate}

\noindent The WBB algorithm is fast and scalable to compute a regularized estimator. For a large number of popular priors, the minimizing solution $\theta^*_{{\bf w},n}$ in the second step can be directly obtained via regularization packages such as {\tt glmnet} by Trevor Hastie and {\tt genlasso} by Taylor Arnold. When the likelihood function or the prior is specially designed, Stochastic Gradient Descent (SGD) is powerful and fast enough to solve the minimization problem. It can be easily implemented in TensorFlow once the objective function is specified. See Appendix (\ref{SGD}) and \cite{polson2017deep} for further discussion. \\

\noindent  The next section builds on \cite{newton1994approximate} and derives asymptotic properties of the weighted Bayesian Bootstrap. We simply add the regularized factor. To choose the amount of regularization $\lambda$, we can use the marginal likelihood $m_\lambda (y)$, estimated by bridge sampling (\cite{gelman1998simulating}) or simply using predictive cross-validation.\\

\subsection{WBB Properties}
\noindent The following proposition which follows from the Theorem 2 in \cite{newton1994approximate} summaries the properties of WBB.\\

\noindent \textbf{Proposition}
{\it The weighted Bayesian Bootstrap draws are approximate posterior samples
$$
\left\{\theta^{*(k)}_{{\bf w},n} \right\}_{k=1}^K \sim p(\theta | y).
$$}

\noindent Now we consider `large $n$' properties. The variation in the posterior density $p(\theta|y) \propto e^{-n l_n(\theta)}p(\theta)$ for sufficiently large $n$ will be dominated by the likelihood term. Expanding $l_n(\theta)$ around its maximum, $\hat\theta$, and defining $J_n(\hat\theta) = nj(\hat\theta)$ as the observed information matrix gives the traditional normal approximation for the posterior distribution
$$
\theta \sim N_d \left( \hat\theta_n, J_n^{-1}(\hat\theta)\right)
$$
where $ \hat{\theta}_n$ is the MLE. A more accurate approximation is obtained by expanding around the posterior mode, $\theta^*$, which we will exploit in our weighted Bayesian Bootstrap. Now we have the asymptotic distributional approximation 
$$
\theta \sim N_d \left( \theta^*, J_n^{-1}(\theta^*)\right)
$$
where $\theta^*_n \defeq {\rm arg \; max}_\theta \; p( \theta|y) $ is the posterior mode. \\

\noindent The use of the posterior mode here is crucially important as it's the mode that is computationally available from TensorFlow and Keras. Approximate normality and second order approximation also holds, see \cite{johnson1970asymptotic}, \cite{Bertail} and \cite{newton1994approximate} for future discussion. Specifically,
$$
\sqrt{ n I ( \hat{\theta}_n )} \left ( {\theta}^*_n - \hat{\theta}_n \right ) \eqdist Z
$$
where $Z \sim N(0,1)$ is a standard Normal variable. The conditional posterior satisfies $\mathbb{P} \left ( | {\theta}^*_n - \hat{\theta}_n | > \epsilon \right ) \rightarrow 0$ for each $ \epsilon > 0 $ as $ n \rightarrow \infty $. In the 'large $p$' case, a number of results are available for posterior concentration, for example, see \cite{van2014horseshoe} for sparse high dimensional models.

\section{Applications}
\noindent Consider now a number of scenarios to assess when WBB corresponds to a full Bayesian posterior distribution. 
\subsection{Lasso}
First, a simple univariate normal means problem with a  lasso prior where  
$$
y|\theta \sim N(\theta,1^2), \quad \theta \sim Laplace (0,1/\lambda)
$$

\begin{figure}[!ht]
\centering 
\includegraphics[scale=0.55]{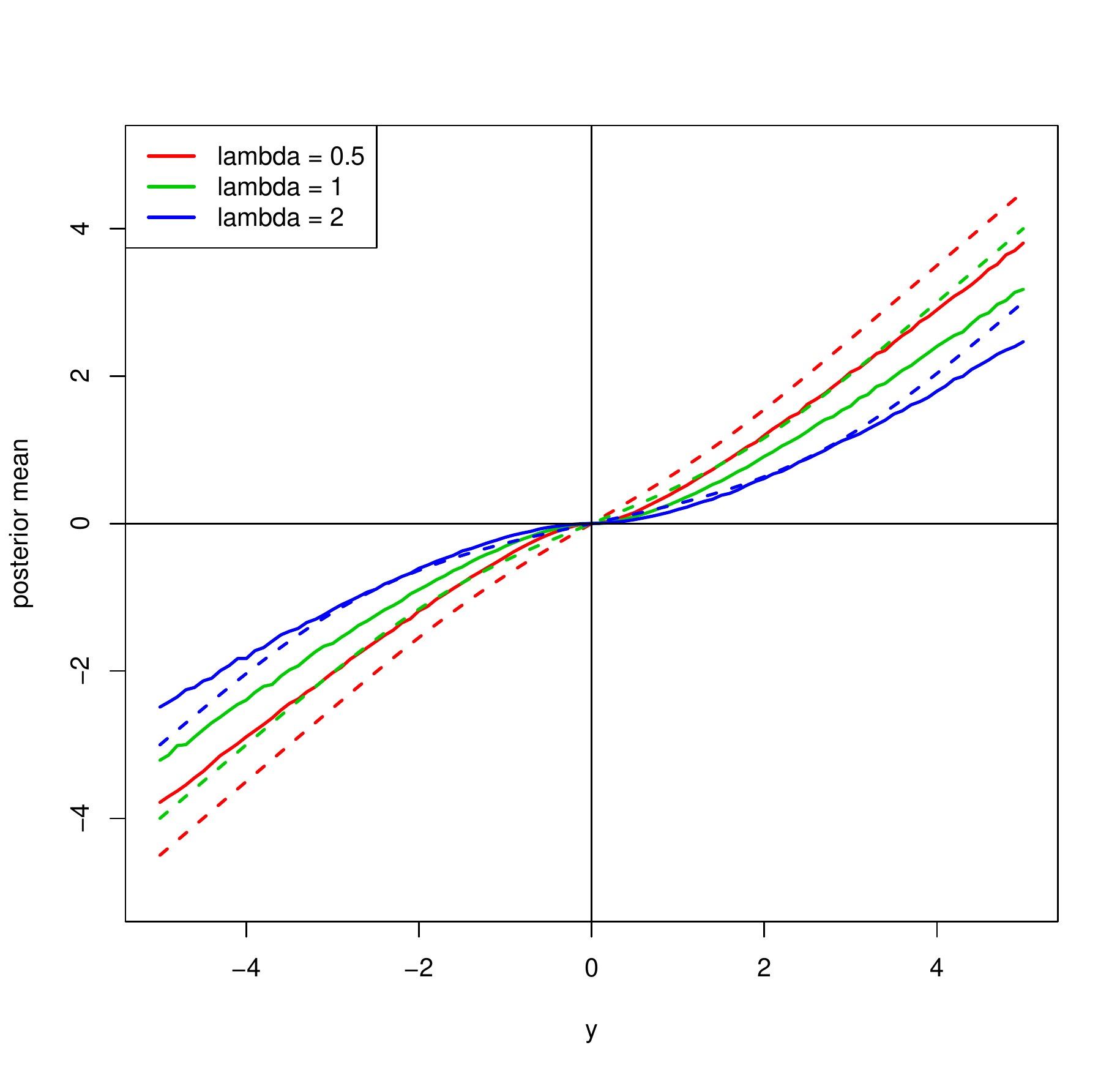} 
\caption{Normal means model with lasso prior: WBB mean $E_{{\bf w}}(\theta^*_{\bf w}|y)$ (in solid lines) versus exact posterior mean $E(\theta|y)$ (in dashed lines).}
\label{simple}
\end{figure}

\noindent Given the i.i.d. exponential weights $w_1$ and $w_2$, the weighted posterior mode $\theta^*_{\bf w}$ is 
$$
\theta^*_{\bf w} = \underset{\theta \in \Theta}{\argmin}\, \left\{\frac{w_1}{2} (y-\theta)^2 + \lambda w_2 |\theta|\right\}.
$$
This is sufficiently simple for an exact WBB solution in terms of soft thresholding:
$$
\theta^*_{\bf w} = 
\begin{cases}
y - \lambda w_2/w_1 &\mbox{if $y > \lambda w_2/w_1$},\\
y + \lambda w_2/w_1 &\mbox{if $y < -\lambda w_2/w_1$},\\
0 &\mbox{if $|y| \leq \lambda w_2/w_1$}.\\
\end{cases}
$$
The WBB mean $E_{{\bf w}}(\theta^*_{\bf w}|y)$ is approximated by the sample mean of $\{ \theta^{*(k)}_{\bf w}\}_{k=1}^K$. On the other hand, \cite{mitchell1994note} gives the expression for the posterior mean, 

\begin{eqnarray*}
E(\theta|y) &=& \frac{\int_{-\infty}^\infty \theta\exp\left\{-(y-\theta)^2/2 - \lambda |\theta|\right\} d\theta}{\int_{-\infty}^\infty \exp\left\{-(y-\theta)^2/2 - \lambda |\theta|\right\} d\theta}\\
&=& \frac{F(y)}{F(y) + F(-y)}(y+\lambda) + \frac{F(-y)}{F(y) + F(-y)}(y-\lambda)\\
&=& y + \frac{F(y) - F(-y)}{F(y) + F(-y)}\lambda
\end{eqnarray*}
where $F(y) = \exp(y)\Phi(-y-\lambda)$ and $\Phi(\cdot)$ is the c.d.f. of standard normal distribution. We plot  the WBB mean versus the exact posterior mean in Figure (\ref{simple}).  Interestingly, WBB algorithm gives sparser posterior means. 

\subsection{Diabetes Data}
To illustrate our methodology, we use weighted Bayesian Bootstrap (WBB) on the classic diabetes dataset. The measurements for 442 diabetes patients are obtained ($n=442$), with 10 baseline variables ($p=10$), such as age, sex, body mass index, average blood pressure, and six blood serum measurements. \\

\noindent The likelihood function is given by
$$
l(y|\beta) = \prod_{i=1}^n p(y_i|\beta)
$$
where
$$
p(y_i|\beta) = \frac{1}{\sqrt{2\pi}\sigma} \exp\left\{-\frac{1}{2\sigma^2}(y_i - x_i'\beta)^2 \right\}.
$$

\noindent We draw 1000 sets of weights ${\bf{w}} = \{w_i\}_{i=1}^{n+1}$ where $w_i$'s are i.i.d. exponentials. For each weight set, the weighted Bayesian estimate $\beta^*_{\bf w}$ is calculated using (\ref{eqn:diabetes}) via the regularization method in the package {\tt glmnet}. 

\begin{equation}
\label{eqn:diabetes}
\hat\beta_{\bf{w}} := \underset{\beta}{\argmin} \;\sum_{i=1}^n w_i(y_i - x_i'\beta)^2+ \lambda w_{n+1}\sum_{j=1}^p |\beta_j|\; . 
\end{equation}
The regularization factor $\lambda$ is chosen by cross-validation with unweighted likelihood. The weighted Bayesian Bootstrap is also performed with fixed prior, namely, $w_{n+1}$ is set to be 1 for all bootstrap samples. \cite{polson2014bayesian} analyze the same dataset using the Bayesian Bridge estimator and suggest MCMC sampling from the posterior. \\

\noindent To compare our WBB results we also run the Bayesian bridge estimation. Here the Bayesian setting we use is 
$$p(\beta, \sigma^2) = p(\beta | \sigma^2)p(\sigma^2),\, \text{where }  p(\sigma^2) \propto 1/\sigma^2.$$
The prior on $\beta$, with suitable normalization constant $C_\alpha$, is given by $$p(\beta) = C_\alpha\exp(-\sum_{j=1}^p |\beta_j/\tau|^\alpha  ).$$ The hyper-parameter is drawn as $\nu = \tau^{-\alpha} \sim \Gamma(2,2)$, where $\alpha = 1/2.$\\

\begin{figure}[!ht]
\centering 
\includegraphics[scale=0.9]{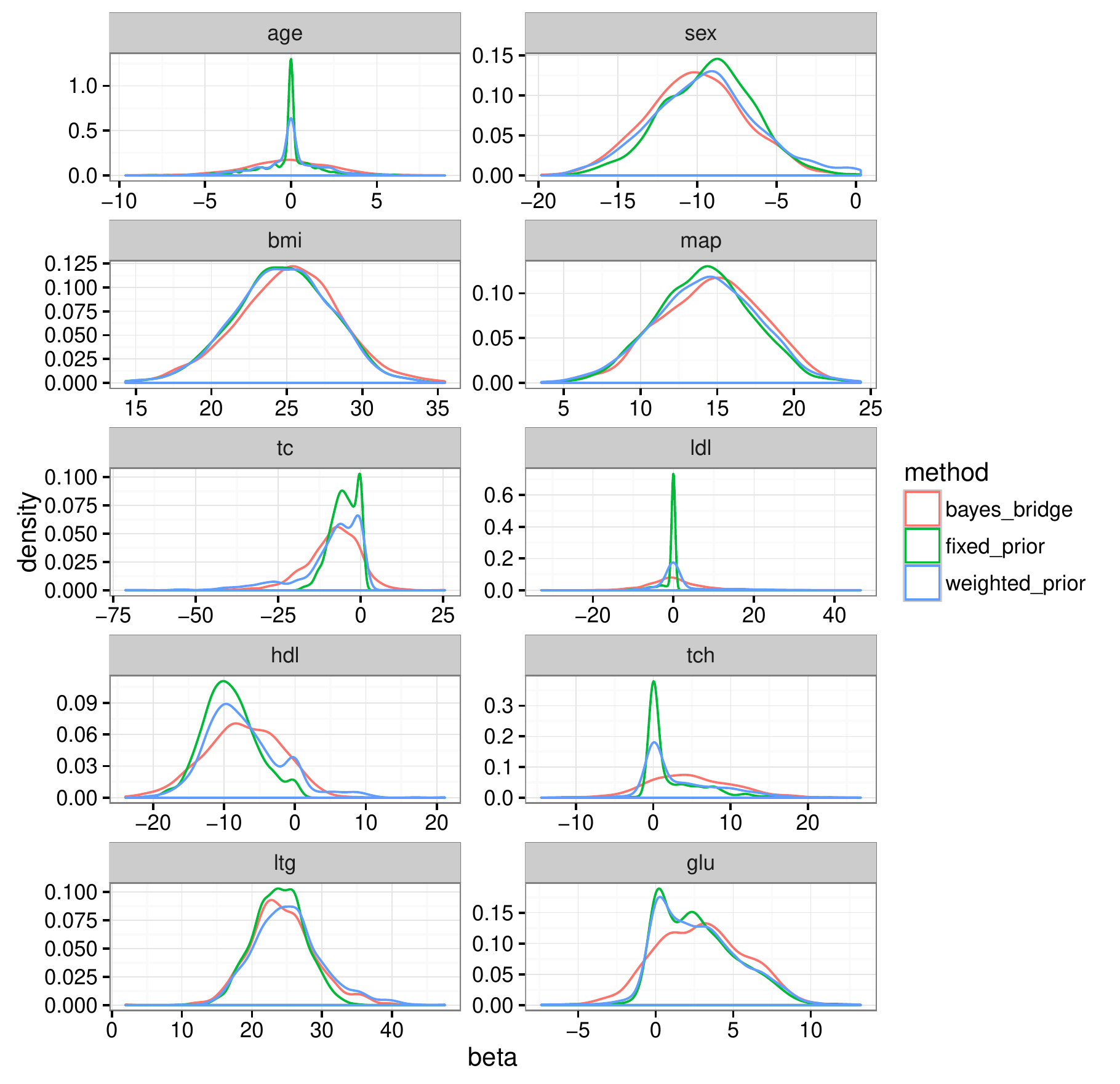} 
\caption{Diabetes example: the weighted Bayesian Bootstrap (with fixed prior and weighted prior) and Bayesian Bridge are used to draw from the marginal posteriors for $\beta_j$'s, j = 1,2,...10. }
\label{diabetes}
\end{figure}

\noindent Figure (\ref{diabetes}) shows the results of all these three methods (the weighted Bayesian Bootstrap with fixed prior / weighted prior and the Bayesian Bridge). Marginal posteriors for $\beta_j$'s are presented. One notable feature is that the weighted Bayesian Bootstrap tends to introduce more sparsity than Bayesian Bridge does. For example, the weighted Bayesian Bootstrap posteriors of {\tt age, ldl} and {\tt tch} have higher spikes located around 0, compared with the Bayesian Bridge ones. For {\tt tc, hdl, tch} and {\tt glu}, multi-modes in the marginal posteriors are observed. In general, the posteriors with fixed priors are more concentrated than those with randomly weighted priors. This difference is naturally attributed to the certainty in the prior weights.

\subsection{Trend Filtering}
The generalized lasso solves the optimization problem:
\begin{eqnarray}
\beta^* &=& \underset{\beta}{\argmin} \, \left\{ l(y|\beta) + \lambda\phi(\beta)  \right\}\\
&=& \underset{\beta}{\argmin} \, \frac{1}{2}\|y - X\beta\|_2^2 + \lambda \|D\beta\|_1
\end{eqnarray}
where $ l(y|\beta) = \frac{1}{2}\|y - X\beta\|_2^2 $ is the negative log-likelihood. $D \in \R^{m\times p}$ is a penalty matrix and $ \lambda\phi(\beta) = \lambda \|D\beta\|_1$ is the negative log-prior or regularization penalty. There are fast path algorithms for solving this problem (see {\tt genlasso} package).\\

\noindent   As a subproblem, polynomial trend filtering (\cite{tibshirani2014adaptive, polson2015mixtures}) is recently introduced for piece-wise polynomial curve-fitting, where the knots and the parameters are chosen adaptively. Intuitively, the trend-filtering estimator is similar to an adaptive spline model: it penalizes the discrete derivative of order $k$, resulting in piecewise polynomials of higher degree for larger $k$.\\

\begin{figure}[!ht]
\centering 
\includegraphics[scale=0.55]{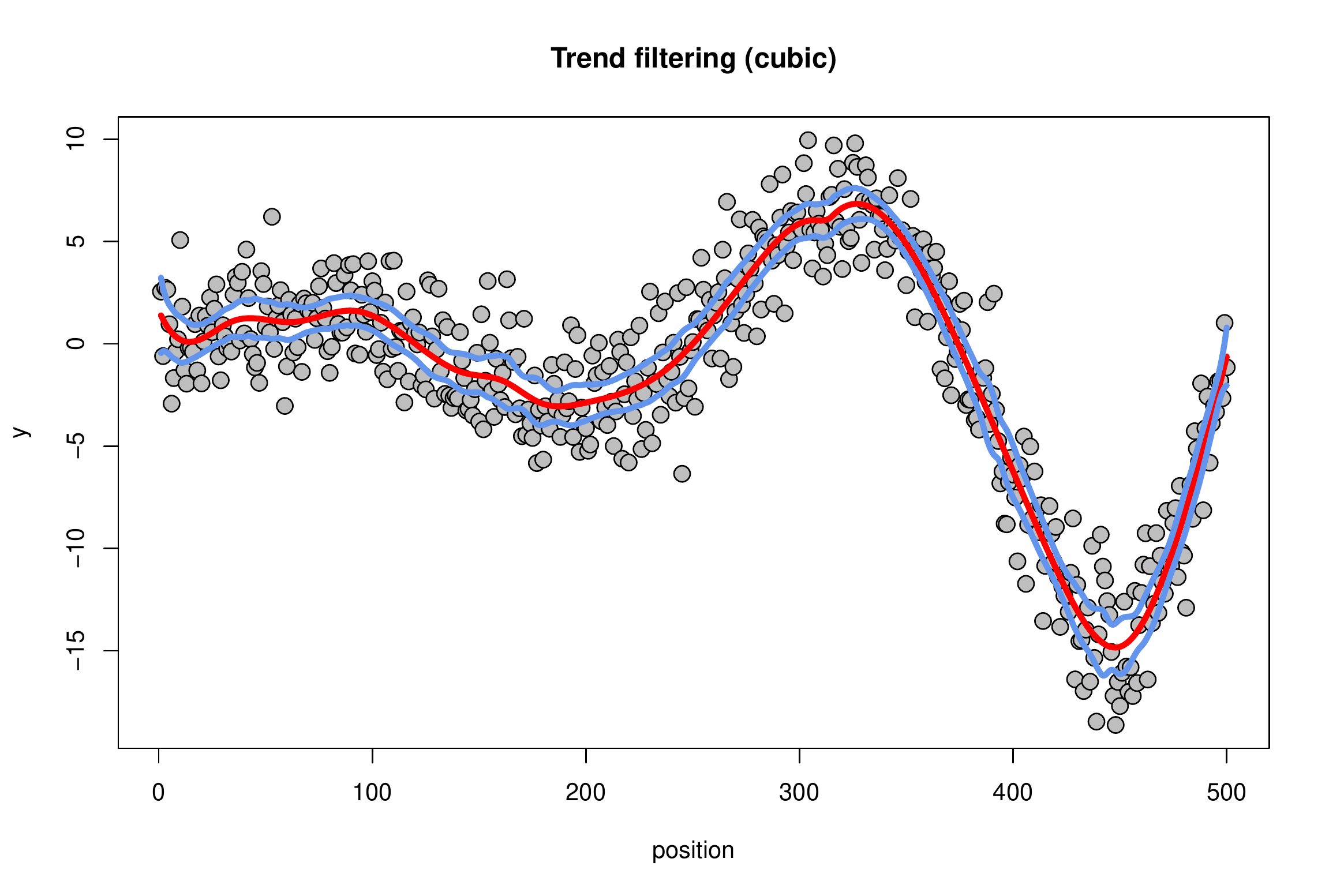} 
\caption{Cubic trend filtering: the red line is $\hat\beta_i$ for $i=1,2,...500$; the blue line is $\hat\beta_i \pm 2*se$ where the standard errors are easily computed by WBB. $\lambda = 1000$.}
\label{filter}
\end{figure}

\noindent Specifically, $X = I_p$ in the trend filtering setting  and the data $y = (y_1, ..., y_p)$ are assumed to be meaningfully ordered from 1 to $p$. The penalty matrix is specially designed by the discrete $(k+1)$-th order  derivative,
$$D^{(1)}=
\begin{bmatrix}
    -1       & 1 & 0 & \dots & 0 & 0 \\
    0       & -1 & 1 & \dots & 0 & 0 \\
    \hdotsfor{5} \\
    0       & 0 & 0 & \dots & -1 & 1
\end{bmatrix}_{(p-1)\times p}
$$
and $D^{(k+1)} = D^{(1)}D^{(k)}$ for $k =1,2,3...$. For example, the log-prior in linear trend filtering is explicitly written as $\lambda\sum_{i=1}^{p-2}|\beta_{i+2} - 2\beta_{i+1} + \beta_{i}|$. For a general order $k >1$, 
$$
\|D^{(k+1)}\beta\|_1 = \sum_{i=1}^{p-k-1} \Big| \sum_{j=i}^{i+k+1} (-1)^{(j-i)} \binom{k+1}{j-i}\beta_j \Big|.
$$
WBB solves the following generalized lasso problem in each draw:
\begin{eqnarray*}
 \beta_{\bf w}^* &=& \underset{\beta}{\argmin} \, \frac{1}{2}\sum_{i=1}^p w_i(y_i - \beta_i)^2 + \lambda w_{p+1}\|D^{(k)}\beta\|_1\\
&=&\underset{\beta}{\argmin} \, \frac{1}{2}  \|Wy - W\beta\|_2^2  + \lambda \|D^{(k)}\beta\|_1\\
&=&W^{-1}  \underset{\tilde\beta}{\argmin} \, \frac{1}{2}  \|\tilde{y}_{\bf w} - \tilde{\beta}_{\bf w}\|_2^2  + \lambda \|\tilde{D}^{(k)}_{\bf w}\tilde{\beta}_{\bf w}\|_1\\
\end{eqnarray*}
where $$W = diag(\sqrt{w_i}/\sqrt{w_{p+1}}, ...,  \sqrt{w_p}/\sqrt{w_{p+1}})$$ and $$\tilde{y}_{\bf w} = Wy, \, \tilde{\beta}_{\bf w} = W\beta, \,   \tilde{D}^{(k)}_{\bf w}= D^{(k)}W^{-1}.$$\\
\noindent To illustrate our method, we simulate data $y_i$ from a Fourier series regression 
$$
y_i = \sin\left(\frac{4\pi}{500} i\right)\exp\left({\frac{3}{500} i}\right) + \epsilon_i
$$
for $i=1,2,...500$, where $\epsilon_i\sim N(0,2^2)$ are i.i.d. Gaussian noises. The cubic trend filtering result is given in Figure (\ref{filter}). \\

\noindent For each $i$, the weighted Bayesian Bootstrap gives a group of estimates $\left\{\beta_{\bf w}^*(i) \right\}_{j=1}^T$ where $T$ is the total number of draws. The standard error of $\hat\beta_i$ is easily computed using these weighted bootstrap estimates.

\subsection{Deep Learning: MNIST Example}
Deep learning is a form of machine learning that uses hierarchical abstract layers of latent variables to perform pattern matching and prediction. \cite{polson2017deep} take a Bayesian probabilistic perspective and provide a number of insights into more efficient algorithms for optimization
and hyper-parameter tuning.\\

\noindent The general goal is to finds a predictor of an output $y$ given a high dimensional input $x$. For a classification problem, $y \in \{1, 2, ..., K\}$ is a discrete variable and can be coded as a $K$-dimensional 0-1 vector. The model is as follows. Let $z^{(l)}$ denote the $l$-th layer, and so $x = z^{(0)}$. The final output is the response $y$,
which can be numeric or categorical. A deep prediction rule is then 

\begin{align*}
z^{(1)} & = f^{(1)} \Big( W^{(0)} x + b^{(0)} \Big),\\
z^{(2)} & = f^{(2)} \Big( W^{(1)} z^{(1)} + b^{(1)} \Big),\\
& \cdots \\
z^{(L)} & = f^{(L)} \Big( W^{(L-1)} z^{(L-1)} + b^{(L-1)} \Big),\\
\hat{y} (x) & = z^{(L)}.
\end{align*}

\noindent Here, $W^{(l)}$ are weight matrices, and $b^{(l)}$ are threshold or activation levels. $f^{(l)}$ is the activation function. Probabilistically, the output $y$ in a classification problem is generated by a probability model 
$$
p(y|x, W, b) \propto \exp\{-l(y|x, W, b)\}
$$
where $l(y|x, W, b) = \sum_{i=1}^{n}l_i(y_i|x_i, W, b) $ is the  negative cross-entropy,
$$
l_i(y_i|x_i, W, b) = l_i(y_i, \hat{y}(x_i)) = \sum_{k=1}^K y_{ik}\log\hat{y}_k(x_i)
$$
where $y_{ik}$ is 0 or 1 and $K = 10$.
Adding the negative log-prior $\lambda\phi(W, b)$, the objective function (negative log-posterior) to be minimized by stochastic gradient descent is 
$$
\mathcal{L}_\lambda(y,\hat{y}) = \sum_{i=1}^{n}l_i(y_i, \hat{y}(x_i)) + \lambda\phi(W, b).
$$
Accordingly, with each draw of weights ${\bf w}$, WBB provides the estimates $(W^*_{\bf w}, b^*_{\bf w})$ by solving the following optimization problem.
$$
(W^*_{\bf w}, b^*_{\bf w}) = \argmin_{W, b} \sum_{i=1}^{n}w_i l_i(y_i|x_i, W, b) + \lambda w_p\phi(W, b)
$$

\noindent We take the classic MNIST example to illustrate the application of WBB in deep learning. The MNIST database of handwritten digits, available from Yann LeCun's website, has 60,000 training examples and 10,000 test examples. Here the high-dimensional $x$ is a normalized and centered fixed-size ($28\times 28$) image and the output $\hat{y}$ is a 10-dimensional vector, where $i$-th coordinate corresponds to the probability of that image being the $i$-th digit.\\

\noindent For simplicity, we build a 2-layer neural network with layer sizes 128 and 64 respectively. Therefore, the dimensions of parameters are
\begin{align*}
W^{(0)} \in \R^{128\times 784},\, b^{(0)}\in \R^{128},\\
W^{(1)} \in \R^{64\times 128},\, b^{(1)}\in \R^{64},\\
W^{(2)} \in \R^{10\times 64},\, b^{(0)}\in \R^{10}.
\end{align*}
The activation function $f^{(i)}$ is ReLU, $f(x) = \max\{0,x\}$, and the negative log-prior is specified as 
$$
\lambda\phi(W,b) = \lambda\sum_{l=0}^{2}\Vert W^{(l)} \Vert_2^2
$$
where $\lambda = 10^{-4}$. \\

\begin{figure}[!ht]
	\centering
	\includegraphics[width=0.7\linewidth]{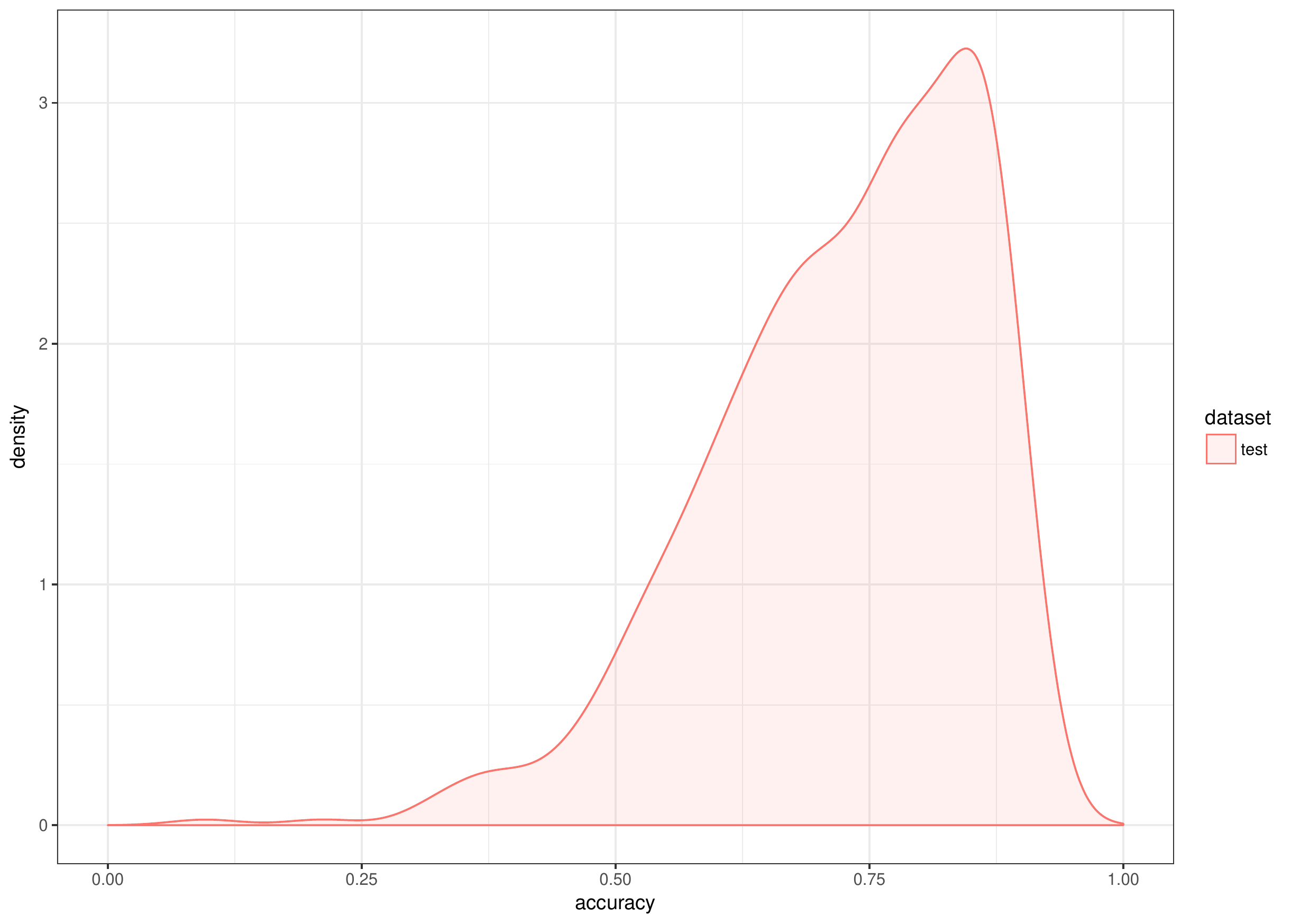}
	\caption{Posterior distribution of the classification accuracy. $n=500, \lambda=10^{-4}$.}
	\label{fig:acc}
\end{figure}

\noindent Figure (\ref{fig:acc}) shows the posterior distribution of the classification accuracy in the test dataset. We see that the test accuracies are centered around 0.75 and the posterior distribution is left-skewed. Furthermore, the accuracy is higher than 0.35 in 99\% of the cases. The 95\% interval is [0.407, 0.893].

\section{Discussion}
Weighted Bayesian Bootstrap (WBB) provides a computationally attractive solution to scalable Bayesian inference (\cite{minsker2014scalable, welling2011bayesian})
whilst accounting for parameter uncertainty by drawing  samples from a weighted posterior distribution. WBB can also be used in conjunction with proximal methods (\cite{parikh2013proximal}, \cite{polson2015proximal})
to provide  sparsity in high dimensional statistica problems. With a similar ease of computation, WBB provides an alternative to ABC methods (\cite{beaumont2009adaptive}) and Variational Bayes (VB) methods. A fruitful area for future research is the comparison of approximate Bayesian computation with simulated Bayesian Bootstrap inference.

\nocite{*}
\bibliography{michael-jianeng.bib}

\begin{thebibliography}{}

\bibitem[\protect\citeauthoryear{Beaumont, Cornuet, Marin, and Robert}{Beaumont
  et~al.}{2009}]{beaumont2009adaptive}
Beaumont, M.~A., J.-M. Cornuet, J.-M. Marin, and C.~P. Robert (2009).
\newblock Adaptive approximate bayesian computation.
\newblock {\em Biometrika\/}~{\em 96\/}(4), 983--990.

\bibitem[\protect\citeauthoryear{Bertail and Lo}{Bertail and
  Lo}{1991}]{Bertail}
Bertail, P. and A.~Y. Lo (1991).
\newblock On johnson's asymptotic expansion for a posterior distribution.
\newblock {\em Centre de Recherche en Economie et Statistique\/}.

\bibitem[\protect\citeauthoryear{Daniels and Young}{Daniels and
  Young}{1991}]{daniels1991saddlepoint}
Daniels, H. and G.~Young (1991).
\newblock Saddlepoint approximation for the studentized mean, with an
  application to the bootstrap.
\newblock {\em Biometrika\/}~{\em 78\/}(1), 169--179.

\bibitem[\protect\citeauthoryear{Efron}{Efron}{1981}]{efron1981nonparametric}
Efron, B. (1981).
\newblock Nonparametric standard errors and confidence intervals.
\newblock {\em Canadian Journal of Statistics\/}~{\em 9\/}(2), 139--158.

\bibitem[\protect\citeauthoryear{Efron}{Efron}{2012}]{efron2012bayesian}
Efron, B. (2012).
\newblock {Bayesian} inference and the parametric bootstrap.
\newblock {\em The Annals of Applied Statistics\/}~{\em 6\/}(4), 1971.

\bibitem[\protect\citeauthoryear{Gelman and Meng}{Gelman and
  Meng}{1998}]{gelman1998simulating}
Gelman, A. and X.-L. Meng (1998).
\newblock Simulating normalizing constants: From importance sampling to bridge
  sampling to path sampling.
\newblock {\em Statistical Science\/}, 163--185.

\bibitem[\protect\citeauthoryear{Gordon, Salmond, and Smith}{Gordon
  et~al.}{1993}]{gordon1993novel}
Gordon, N.~J., D.~J. Salmond, and A.~F. Smith (1993).
\newblock Novel approach to nonlinear/non-gaussian {Bayesian} state estimation.
\newblock In {\em IEE Proceedings F-Radar and Signal Processing}, Volume 140,
  pp.\  107--113. IET.

\bibitem[\protect\citeauthoryear{Gramacy and Polson}{Gramacy and
  Polson}{2012}]{gramacy2012simulation}
Gramacy, R.~B. and N.~G. Polson (2012).
\newblock Simulation-based regularized logistic regression.
\newblock {\em Bayesian Analysis\/}~{\em 7\/}(3), 567--590.

\bibitem[\protect\citeauthoryear{Hans}{Hans}{2009}]{hans2009Bayesian}
Hans, C. (2009).
\newblock {Bayesian} lasso regression.
\newblock {\em Biometrika\/}~{\em 96\/}(4), 835--845.

\bibitem[\protect\citeauthoryear{Johnson}{Johnson}{1970}]{johnson1970asymptotic}
Johnson, R.~A. (1970).
\newblock Asymptotic expansions associated with posterior distributions.
\newblock {\em The annals of mathematical statistics\/}~{\em 41\/}(3),
  851--864.

\bibitem[\protect\citeauthoryear{Lopes, Polson, and Carvalho}{Lopes
  et~al.}{2012}]{lopes2012bayesian}
Lopes, H.~F., N.~G. Polson, and C.~M. Carvalho (2012).
\newblock Bayesian statistics with a smile: a resampling-sampling perspective.
\newblock {\em Brazilian Journal of Probability and Statistics\/}, 358--371.

\bibitem[\protect\citeauthoryear{Minsker, Srivastava, Lin, and Dunson}{Minsker
  et~al.}{2014}]{minsker2014scalable}
Minsker, S., S.~Srivastava, L.~Lin, and D.~Dunson (2014).
\newblock Scalable and robust bayesian inference via the median posterior.
\newblock In {\em International Conference on Machine Learning}, pp.\
  1656--1664.

\bibitem[\protect\citeauthoryear{Mitchell}{Mitchell}{1994}]{mitchell1994note}
Mitchell, A.~F. (1994).
\newblock A note on posterior moments for a normal mean with double-exponential
  prior.
\newblock {\em Journal of the Royal Statistical Society. Series B
  (Methodological)\/}, 605--610.

\bibitem[\protect\citeauthoryear{Newton}{Newton}{1991}]{newton1991weighted}
Newton, M.~A. (1991).
\newblock {\em The weighted likelihood bootstrap and an algorithm for
  prepivoting}.
\newblock Ph.\ D. thesis, Department of Statistics, University of Washington,
  Seattle.

\bibitem[\protect\citeauthoryear{Newton and Raftery}{Newton and
  Raftery}{1994}]{newton1994approximate}
Newton, M.~A. and A.~E. Raftery (1994).
\newblock Approximate {Bayesian} inference with the weighted likelihood
  bootstrap.
\newblock {\em Journal of the Royal Statistical Society. Series B
  (Methodological)\/}, 3--48.

\bibitem[\protect\citeauthoryear{Parikh and Boyd}{Parikh and
  Boyd}{2013}]{parikh2013proximal}
Parikh, N. and S.~Boyd (2013).
\newblock Proximal algorithms, in foundations and trends in optimization.

\bibitem[\protect\citeauthoryear{Polson and Scott}{Polson and
  Scott}{2015}]{polson2015mixtures}
Polson, N.~G. and J.~G. Scott (2015).
\newblock Mixtures, envelopes and hierarchical duality.
\newblock {\em Journal of the Royal Statistical Society: Series B (Statistical
  Methodology)\/}.

\bibitem[\protect\citeauthoryear{Polson, Scott, and Willard}{Polson
  et~al.}{2015}]{polson2015proximal}
Polson, N.~G., J.~G. Scott, and B.~T. Willard (2015).
\newblock Proximal algorithms in statistics and machine learning.
\newblock {\em Statistical Science\/}~{\em 30\/}(4), 559--581.

\bibitem[\protect\citeauthoryear{Polson, Scott, and Windle}{Polson
  et~al.}{2014}]{polson2014bayesian}
Polson, N.~G., J.~G. Scott, and J.~Windle (2014).
\newblock The {Bayesian} bridge.
\newblock {\em Journal of the Royal Statistical Society: Series B (Statistical
  Methodology)\/}~{\em 76\/}(4), 713--733.

\bibitem[\protect\citeauthoryear{Polson and Sokolov}{Polson and
  Sokolov}{2017}]{polson2017deep}
Polson, N.~G. and V.~Sokolov (2017).
\newblock Deep learning: A bayesian perspective.
\newblock {\em Bayesian Analysis\/}~{\em 12\/}(4), 1275--1304.

\bibitem[\protect\citeauthoryear{Rubin}{Rubin}{1981}]{rubin1981Bayesian}
Rubin, D.~B. (1981).
\newblock The {Bayesian} bootstrap.
\newblock {\em The Annals of Statistics\/}~{\em 9\/}(1), 130--134.

\bibitem[\protect\citeauthoryear{Strawderman, Wells, and Schifano}{Strawderman
  et~al.}{2013}]{strawderman2013hierarchical}
Strawderman, R.~L., M.~T. Wells, and E.~D. Schifano (2013).
\newblock Hierarchical bayes, maximum a posteriori estimators, and minimax
  concave penalized likelihood estimation.
\newblock {\em Electronic Journal of Statistics\/}~{\em 7}, 973--990.

\bibitem[\protect\citeauthoryear{Taddy, Chen, Yu, and Wyle}{Taddy
  et~al.}{2015}]{taddy2015bayesian}
Taddy, M., C.-S. Chen, J.~Yu, and M.~Wyle (2015).
\newblock Bayesian and empirical bayesian forests.
\newblock {\em arXiv preprint arXiv:1502.02312\/}.

\bibitem[\protect\citeauthoryear{Tibshirani}{Tibshirani}{2014}]{tibshirani2014adaptive}
Tibshirani, R.~J. (2014).
\newblock Adaptive piecewise polynomial estimation via trend filtering.
\newblock {\em The Annals of Statistics\/}~{\em 42\/}(1), 285--323.

\bibitem[\protect\citeauthoryear{Van Der~Pas, Kleijn, and Van Der~Vaart}{Van
  Der~Pas et~al.}{2014}]{van2014horseshoe}
Van Der~Pas, S., B.~Kleijn, and A.~Van Der~Vaart (2014).
\newblock The horseshoe estimator: Posterior concentration around nearly black
  vectors.
\newblock {\em Electronic Journal of Statistics\/}~{\em 8\/}(2), 2585--2618.

\bibitem[\protect\citeauthoryear{Welling and Teh}{Welling and
  Teh}{2011}]{welling2011bayesian}
Welling, M. and Y.~W. Teh (2011).
\newblock Bayesian learning via stochastic gradient langevin dynamics.
\newblock In {\em Proceedings of the 28th International Conference on Machine
  Learning (ICML-11)}, pp.\  681--688.

\bibitem[\protect\citeauthoryear{West}{West}{1992}]{west1992modelling}
West, M. (1992).
\newblock Modelling with {Mixtures}.
\newblock {\em In Bayesian Statistics 4\/}, 503--524.

\end{thebibliography}

\appendix
\section{Stochastic Gradient Descent (SGD) \label{SGD}}
Stochastic gradient descent (SGD) method or its variation  is typically used to find the deep learning model weights by minimizing the penalized loss function, $\sum_{i=1}^n w_i l_i(y_i; \theta) + \lambda w_{p} \phi(\theta)$. The method minimizes the function by taking a negative step along an estimate $g^k$ of the gradient $\nabla\left[\sum_{i=1}^n w_i l_i(y_i; \theta^k) + \lambda w_{p} \phi(\theta^k)\right] $ at iteration $k$. 
The approximate gradient is estimated by calculating 
\[
g^k = \frac{n}{b_k} \sum_{i \in E_k} w_i\nabla  l_i(y_i; \theta^k) +  \lambda w_{p} \frac{n}{b_k}\nabla \phi(\theta^k)
\]
Where $E_k \subset \{1,\ldots,n \}$ and $b_k = |E_k|$ is the number of elements in $E_k$. When $b_k >1$ the algorithm is called batch SGD and simply SGD otherwise. A usual strategy to choose subset $E$ is to go cyclically and pick consecutive elements of $\{1,\ldots,T \}$, $E_{k+1} = [E_k \mod n]+1$. The approximated direction  $g^k$ is calculated using a chain rule (aka back-propagation) for deep learning. It  is an unbiased estimator. Thus, at each iteration, the SGD updates the solution
\[
\theta^{k+1} = \theta^k - t_k g^k
\]
For deep learning applications the step size $t_k$ (a.k.a learning rate) is usually kept constant or some simple step size reduction strategy is used, $t_k = a\exp(-kt)$. Appropriate learning rates or the hyperparameters of reduction schedule  are usually found empirically from numerical experiments and observations of the loss function progression. \\

\end{document}